\documentclass[%
 reprint,
superscriptaddress,
 amsmath,amssymb,
aps,
]{revtex4-2}

\usepackage{graphicx}
\usepackage{dcolumn}
\usepackage{bm}

\DeclareMathAlphabet      {\mathbfit}{OML}{cmm}{b}{it}
\newcommand{\sslash}{\mathbin{/\mkern-5mu/}}

\begin{document}

\preprint{APS/123-QED}

\title{Bidirectional allostery mechanism of catch-bond effect in cell adhesion}

\author{Xingyue Guan}%
\affiliation{%
 Department of Physics, National Laboratory of Solid State Microstructure, and Collaborative Innovation Center of Advanced Microstructures, Nanjing University, Nanjing, Jiangsu 210093, China; 
}%
\affiliation{%
 Wenzhou Key Laboratory of Biophysics, Wenzhou Institute, University of Chinese Academy of Sciences, Wenzhou, Zhejiang 325000, China;
}%

\author{Yunqiang Bian}%
\affiliation{%
 Wenzhou Key Laboratory of Biophysics, Wenzhou Institute, University of Chinese Academy of Sciences, Wenzhou, Zhejiang 325000, China;
}%
\author{Yi Cao}%
\affiliation{%
 Department of Physics, National Laboratory of Solid State Microstructure, and Collaborative Innovation Center of Advanced Microstructures, Nanjing University, Nanjing, Jiangsu 210093, China; 
}%

\author{Wenfei Li}%
 \email{wfli@nju.edu.cn}
\affiliation{%
 Department of Physics, National Laboratory of Solid State Microstructure, and Collaborative Innovation Center of Advanced Microstructures, Nanjing University, Nanjing, Jiangsu 210093, China; 
}%
\affiliation{%
 Wenzhou Key Laboratory of Biophysics, Wenzhou Institute, University of Chinese Academy of Sciences, Wenzhou, Zhejiang 325000, China;
}%

\author{Wei Wang}%
 \email{wangwei@nju.edu.cn}
\affiliation{%
 Department of Physics, National Laboratory of Solid State Microstructure, and Collaborative Innovation Center of Advanced Microstructures, Nanjing University, Nanjing, Jiangsu 210093, China; 
}%

\date{\today}

\begin{abstract}
Catch-bonds, whereby noncovalent ligand-receptor interactions are counterintuitively reinforced by tensile forces, play a major role in cell adhesion under mechanical stress. A basic prerequisite for catch-bond formation is that force-induced remodeling of ligand binding interface occurs prior to bond rupture. However, what strategy receptor proteins utilize to meet such specific kinetic control is still unclear, rendering the mechanistic understanding of catch-bond an open question. Here we report a bidirectional allostery mechanism of catch-bond for the hyaluronan (HA) receptor CD44 which is responsible for rolling adhesion of lymphocytes and circulating tumor cells. Binding of ligand HA allosterically reduces the threshold force for unlocking of otherwise stably folded force-sensing element (i.e., forward allostery), so that much smaller tensile force can trigger the conformational switching of receptor protein to high binding-strength state via backward allosteric coupling before bond rupture. The effect of forward allostery was further supported by performing atomistic molecular dynamics simulations. Such bidirectional allostery mechanism fulfills the specific kinetic control required by catch-bond and is likely to be commonly utilized in cell adhesion. We also revealed a slip-catch-slip triphasic pattern in force response of CD44-HA bond arising from force-induced repartitioning of parallel dissociation pathways. The essential thermodynamic and kinetic features of receptor proteins for shaping the catch-bond were identified.
\end{abstract}

\maketitle
\section{Introduction}\label{sec1}
Cell adhesion plays crucial roles in many biological processes, including bacterial infection, immunity, and tumor metastasis \cite{mcever2010rolling,khalili2015review,janiszewska2020cell,evans2007forces,mathelie2020force,zhu2019mechanosensing,wu2019mechano,limozin2019tcr,kale2018distinct,ishiyama2018force}. In many cases, the receptor-ligand bonds mediating cell adhesion inevitably experience a variety of tensile forces arising from hydrodynamic forces in blood vessels, cytoskeletal motor activity, and deformation of the cell matrix. Generally, the lifetime of the receptor-ligand bond exponentially decreases following the Bell law \cite{bell1978models}, and therefore adhesion becomes weaker, with increasing tensile forces. How cells achieve efficient adhesion under tensile forces is a fundamental question attracting extensive investigation. During the past decades, a growing number of experiments have demonstrated that cells tackle this issue by developing the catch-bond effect, whereby the lifetime of the receptor-ligand bond becomes prolonged after applying appropriate tensile forces \cite{marshall2003direct,thomas2006catch,thomas2008biophysics,thomas2008catch,sokurenko2008catch,sauer2016catch,huang2017vinculin,nord2017catch,puri2019dynein,xie2020dynamics,liu2020high,christophis2011shear}. In addition to cell adhesion, catch bonds are also involved in many other key biomolecular processes, such as fibrin formation in blood clotting \cite{litvinov2018regulatory}, bacterial flagellar motor motion \cite{nord2017catch}, actin depolymerization \cite{lee2013actin}, and even protein folding under force \cite{guo2020hidden}. However, the mechanistic understanding of this counterintuitive catch-bond effect is still far from complete and under debate.

At the molecular level, cell adhesion is mainly mediated by the noncovalent interactions between receptor proteins expressed on the cell surface and ligand/protein molecules on the surface of another cell or the extracellular matrix (Fig. 1a). Several possible mechanisms have been proposed to explain the observed catch-bond effect \cite{thomas2006catch,pereverzev2005two,prezhdo2009theoretical,evans2004mechanical,guo2019understanding,dansuk2019simple,barsegov2005dynamics}, including the allosteric mechanism, two-pathway mechanism, sliding-rebinding mechanism, and other mechanisms. Experimental studies have shown that the catch-bond effect often relies on the conformational plasticity of receptor proteins. For example, adhesion of the $Escherichia$ $coli$ FimH adhesin to the mannose of epithelial cells involves a transition between the closed conformation with high ligand affinity and the open conformation with low ligand affinity \cite{sauer2016catch}. Similarly, selectin-mediated leukocyte rolling adhesion under flow conditions involves transitions between the bent and extended structures \cite{waldron2009transmission,somers2000insights,phan2006remodeling}. More recent experimental work demonstrated the importance of the two-state conformational equilibrium of the hyaluronan (HA) receptor CD44 in mediating rolling adhesion of leukocytes and other cells \cite{suzuki2015mechanical,ogino2010two}. One key step implied in all the classic catch-bond models for these adhesion proteins is the force-induced conformational switching of the ligand binding interface to the high binding-strength state before bond rupture, either directly or allosterically, which prolongs the bond lifetime. However, considering that a tensile force also has the effect of promoting bond rupture by lowering the barrier height before conformational switching to high binding-strength state, one key question that arises and remains unresolved is what strategy receptor proteins use to enable conformational switching prior to bond rupture. To understand the molecular mechanism of such specific kinetic control required by the catch-bond formation, characterizing the interplay between ligand binding, protein allostery, and force-sensing elements is essential. Experimentally investigating such interplay at the single-molecule level is extremely challenging. Molecular dynamics (MD) simulation provides an alternative way to characterize the complex dynamics of biomolecules and has shown great success in elucidating the molecular mechanism of the catch-bond effect \cite{lou2007structure,gunnerson2009atomistic}. However, conventional all-atom MD simulations often encounter sampling difficulties and cannot directly access the timescales of the full molecular events involved in force-regulated cell adhesion. To date, microscopic MD simulations directly reproducing the experimentally observed catch-bond phase in the force-response profile are still lacking.

\begin{figure*}[hbt!]
\centering
\includegraphics[width=0.9\linewidth]{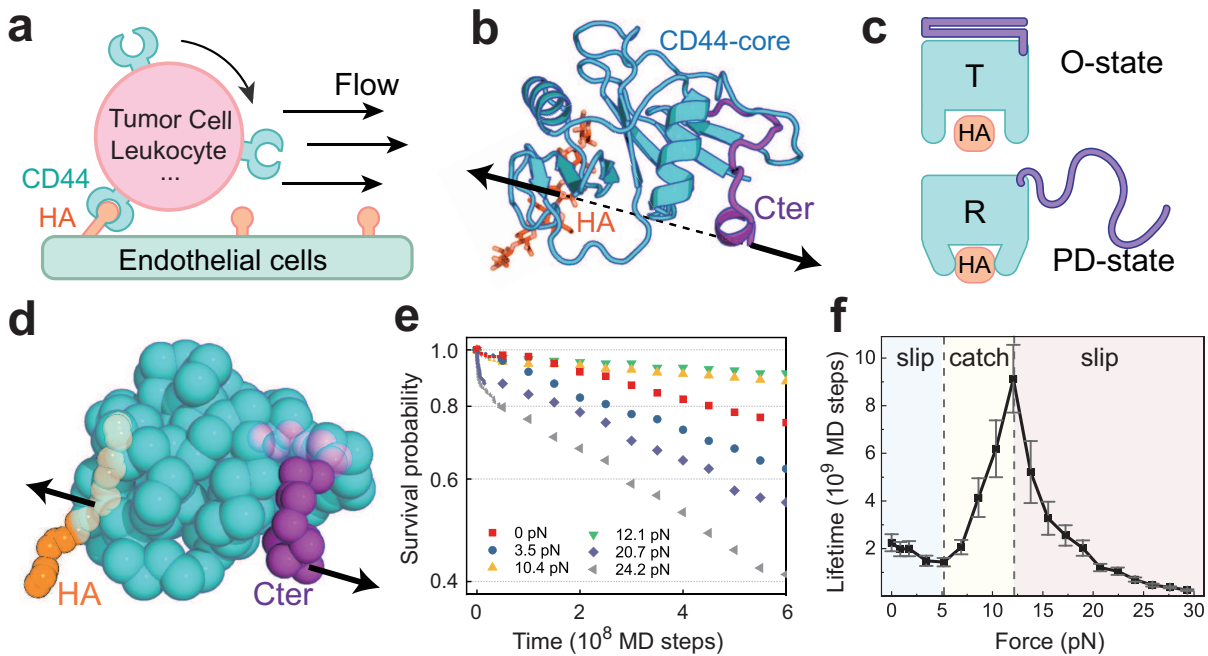}
\caption{Catch-bond effect of CD44-mediated cell adhesion. (a) Cartoon diagram showing the role of the CD44-HA interaction in the rolling adhesion of cells along the blood vessel wall under flow conditions. (b) Three-dimensional structure of CD44 in the ordered state (PDB entry:1UUH\cite{teriete2004structure}). Only the hyaluronan-binding domain of CD44 is shown. The ligand HA is shown by stick representation and its location is reconstructed from the ligand-bound structure (Appendix A). The directions of the tensile force applied during the pulling simulations are labeled by arrows. The HA and Cter were colored by orange and purple, respectively.(c)Cartoon diagrams illustrating the structural features of the ordered (O) and partially disordered (PD) states of CD44. (d) Coarse-grained structure of the CD44-HA system used in the molecular simulations with a residue-resolved dynamic energy landscape model. (e) Survival probability of the CD44-HA bond as a function of simulation time under different tensile forces. (f) Average lifetime of the CD44-HA bond as a function of the applied tensile force. The error bars were estimated by the bootstrap method from at least 192 independent MD simulations.}
\label{fig1}
\end{figure*}

In this work, we develop a residue-resolved dynamic energy landscape model of cell adhesion and conduct MD simulations to study the molecular mechanism of the catch-bond effect by using the HA receptor protein CD44 as a model system. CD44 is a transmembrane receptor mediating the adhesion of a number of cells to the endothelium by specifically binding with the HA ligand (Fig. 1a). It has a functional domain responsible for HA binding, which can interconvert between an ordered (O) structure and a partially disordered (PD) structure (Fig. 1b,c) \cite{takeda2006ligand}. In the ordered state, the C-terminal segment (Cter) connecting to the plasma membrane is well folded, and the core part (CD44-core) adopts a conformation with low HA binding-strength (hereafter termed the tensed (T) conformation). In contrast, in the partially disordered state, the CD44-core adopts a high binding-strength conformation (termed the relaxed (R) conformation), and Cter is disordered. CD44-mediated cell adhesion is essential for various physiological and pathological processes, such as lymphocyte rolling along the blood vessel wall, tumor invasion and metastasis, and development of atherosclerosis \cite{hanke2017cd44}. For these processes, the trafficking of cells expressing CD44 often occurs under flow conditions; therefore, the adhesion of the cells encounters hydrodynamic forces. The rolling adhesion of these cells has been demonstrated to become more efficient when the tensile force exceeds a certain threshold, which was ascribed to the catch-bond effect of CD44-HA binding \cite{suzuki2015mechanical,bano2016single,hanke2017cd44}. 

By performing MD simulations with the dynamic energy landscape model, we not only well reproduced the experimentally observed catch-bond effect of the CD44-HA interactions but also identified a slip-catch-slip triphasic pattern of the force response behavior. Particularly, we revealed a bidirectional allosteric coupling mechanism of the catch-bond effect, whereby the binding of the ligand HA at the initial stage of cell adhesion significantly lowers the threshold force for unlocking of the otherwise stably folded force-sensing C-terminal latch via allosteric coupling, which in turn triggers conformational switching of the CD44-core to the high binding-strength state before bond rupture via backward allosteric coupling. Ligand binding, protein allostery, and force-sensing elements are coordinated by such bidirectional allostery to meet the specific kinetic control required for catch-bond formation.
We also identified the crucial role of the force-induced repartitioning of multiple dissociation pathways and the appropriate conformational equilibrium of the CD44 receptor in shaping the unusual force-response profile of the CD44-HA bond. These findings revealed mechanistic insights into the general physical principle utilized by cells to achieve efficient adhesion under mechanical stress.

\section{Molecular model of receptor-ligand adhesion bond}\label{sec2}
In order to characterize the interplay between ligand binding, protein allostery, and force action involved in catch-bond formation, we constructed a dynamic energy landscape model. In this model, each amino acid of the receptor protein was simplified as a single particle centered on its C$_\alpha$ position, and each disaccharide unit of ligand was represented by five particles (Fig. 1d and Fig. S1 in Supplementary Materials \cite{guan2023SI}). The energy function dictating the protein motions and protein-ligand interactions is given by 
\begin{equation}
V(\overrightarrow{\mathbfit{x}})=V_{apo}\left(\overrightarrow{\mathbfit{x}}\right)+V_{bind}\left(\overrightarrow{\mathbfit{x}}\right)
\end{equation}
Here, $\overrightarrow{\mathbfit{x}}$ collectively represents the coordinates of the residues of receptor protein and ligand beads. The $V_{apo}(\overrightarrow{\mathbfit{x}})$ is the energy function of the receptor protein at apo state. The $V_{bind}\left(\overrightarrow{\mathbfit{x}}\right)$ is the energy function describing the liand binding. As ligand binding reshapes the intrinsic energy landscape, the total energy function dictating the conformational motions of the receptor protein depends on the ligand binding state, which therefore corresponds
to a dynamic energy landscape model. 

For protein CD44, the $V_{apo}(\overrightarrow{\mathbfit{x}})$ energy function can be further decomposed into two parts, dictating the motions of the CD44-core and the folding/unfolding of the force-sensing element Cter, respectively. We used a energy function with a double-basin topography to describe the conformational switch of the CD44-core between the T and R states. The double-basin energy function is given as following \cite{okazaki2006multiple}  
\begin{equation}
\begin{aligned}
V_{DB}&\left(\overrightarrow{\mathbfit{x}}\right) =\left[V\left(\overrightarrow{\mathbfit{x}}\mid\overrightarrow{\mathbfit{x}}^{R}_0\right)+V\left(\overrightarrow{\mathbfit{x}}\mid\overrightarrow{\mathbfit{x}}^{T}_0\right)+\Delta V\right]/2-\\
&\sqrt{\left[\left(V\left(\overrightarrow{\mathbfit{x}}\mid\overrightarrow{\mathbfit{x}}^{R}_0\right)-V\left(\overrightarrow{\mathbfit{x}}\mid\overrightarrow{\mathbfit{x}}^{T}_0\right)-\Delta V\right)/2\right]^2+\Delta^2}
\end{aligned}
\end{equation}
where $V(\overrightarrow{\mathbfit{x}}\mid\overrightarrow{\mathbfit{x}}^{T}_0)$ and $V(\overrightarrow{\mathbfit{x}}\mid\overrightarrow{\mathbfit{x}}^{R}_0)$ are the structure-based potentials centric to the R and T structures of the CD44-core, respectively \cite{li2011frustration,li2014energy,li2012energy}. $\overrightarrow{\mathbfit{x}}^{R}_0$ and $\overrightarrow{\mathbfit{x}}^{T}_0$ are the corresponding coordinates of the coarse-grained beads in the native structures. The key parameter $\Delta V$ controls the relative stabilities (free energy difference) of the T and R conformations, and its value was set according to available experimental restraints \cite{ogino2010two,suzuki2015mechanical}. Details of the dynamic energy landscape model are given in Appendix A and Supplementary Materials \cite{guan2023SI}. The above dynamic energy landscape model provides an explicit description of the ligand binding and mechanical force induced allosteric motions of the receptor protein, by which we can simulate the full molecular events involved in CD44-mediated cell adhesion under tensile forces with molecular dynamics and elucidate the underlying molecular mechanism of the catch-bond effect (Appendix B).

\section{Results}\label{sec2}
\subsection{Slip-catch-slip triphasic pattern in force response of receptor-ligand adhesion bond}
We conducted MD simulations starting from the HA-bound T state (low binding-strength) with a tensile force applied between the Cter of CD44 and the HA ligand. The CD44-HA bond was considered ruptured if the binding energy became zero. By repeating the simulations independently, we calculated the lifetime and survival probability of the CD44-HA complex (Appendix B and Supplementary Text). At zero tensile force, one can observe successful rupture of CD44-HA bond for $\sim$60$\%$ of the MD trajectories within the simulation length(Fig. S2). The distribution of survival probability of CD44-HA bond shows an exponential-like decay feature (Fig. 1e, red). The overall average lifetime was estimated to be $\sim2.2\times10^9$ MD steps based on a maximum likelihood approach (Fig. 1f). This lifetime value, after calibrated by protein conformational dynamics (Supplementary Text) \cite{guan2023SI}, is close to the experimentally measured range of the CD44-HA bond (200-250s) at the force-free case \cite{raman2012distinct}. In comparison, much fewer rupture events were observed at moderate tensile forces, which leads to long-lived CD44-HA bond. For example, with the tensile force of $\sim$10.4pN, only $\sim$20$\%$ of the  MD trajectories show rupture events and the lifetime distribution becomes wider compared to that of the force-free case (Fig. S2). In addition to the prominent distributions at small lifetime range, the events with much longer lifetime have also been significantly populated, and the overall average lifetime was estimated to be $\sim6.2\times10^9$ MD steps(Fig. 1f). Such results suggest that applying appropriate tensile force alters the CD44-HA interactions, leading to observations of more stabilized CD44-HA bond. 

With increasing tensile force, both the survival probability and average lifetime of the CD44-HA bond show intriguing nonmonotonic changes (Fig. 1e, f). In the small tensile force range (0-5.2 pN), the lifetime decreases with tensile force, while the survival probability decays more rapidly. Such results are in accordance with the conventional slip-bond scenario as described by Bell law \cite{bell1978models}. When the tensile force exceeds a certain threshold ($\sim$5.2 pN), the lifetime and survival probability of the CD44 bond increase with the applied tensile force, demonstrating a typical catch-bond effect. With the further increasing of the tensile force ($>$12.1 pN), the lifetime starts to decrease, recovering to slip-bond behavior (Fig. 1f). Overall, the above results not only reproduce the catch-bond effect observed in experiments \cite{christophis2011shear,hanke2017cd44,suzuki2015mechanical} but also identify a previously unrecognized slip-catch-slip triphasic pattern for the force responses of the receptor-HA bond. The success of the dynamic energy landscape model in describing the experimentally observed catch-bond effect enables in-depth investigation of the underlying molecular mechanism.

\begin{figure}[hbt!]
\centering
\includegraphics[width=1.0\linewidth]{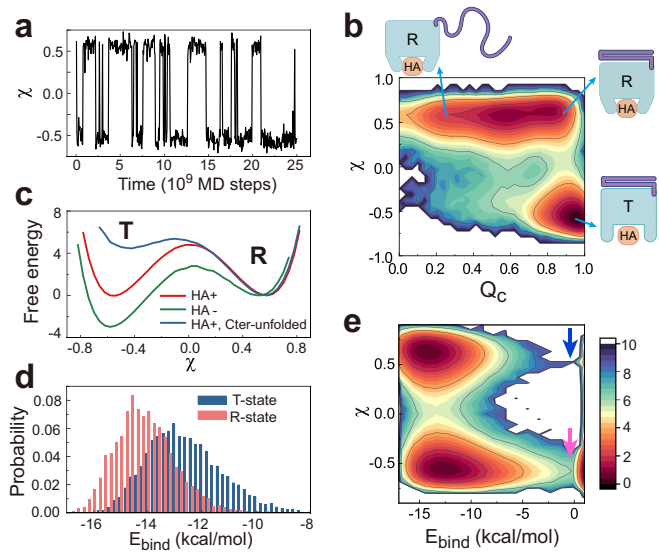}
\caption{Interplay between protein allostery, Cter unfolding, and HA dissociation. (a) Reaction coordinate $\chi$ as a function of time with HA bound. Negative and positive values of $\chi$ correspond to the T and R conformations, respectively. (b) Two-dimensional free energy profile along the reaction coordinates $\chi$ and $Q_C$. Here, $Q_C$ represents the fraction of the formed native contacts between the CD44-core and Cter. The cartoon diagrams for three represent structures were shown schematically. (c) One-dimensional free energy profile along the reaction coordinate $\chi$ demonstrating the relative stabilities of the T and R conformations of the CD44-core with HA unbound (green), with HA bound (red), and with HA bound and Cter unfolded (blue). (d) Distribution of the HA binding energy with the CD44-core staying in the T (blue) and R (red) conformations.(e) Two-dimensional free energy profile along the reaction coordinates $\chi$ and $E_{bind}$. The unit of free energy is kcal/mol.}
\label{fig2}
\end{figure}

\subsection{Allosteric switch to high-affinity state by force-induced unlocking of force-sensing element}
To reveal the underlying molecular mechanism of the observed enhancement of the CD44-HA bond after applying tensile forces, we investigated the conformational dynamics of the HA-bound CD44-core. The trajectories at zero force load show that the CD44-core hops between two conformational states, i.e., the T state ($\chi<0$) and R state ($\chi>0$) (Fig. 2a; see Appendix C for the definitions of reaction coordinates). The two-dimensional free energy profile along the reaction coordinates $Q_{C}$ (which ranges from 0 to 1 and describes the folding event of Cter) and $\chi$ (which describes the conformatonal change of CD44-core) shows that the relative stabilities of the two conformational states can be modulated by the structure of Cter (Fig. 2b). When Cter is largely folded (larger $Q_{C}$), both the T and R states can be significantly populated. In contrast, the CD44-core dominantly stays in the high binding-strength R state after Cter unfolds and detaches from the CD44-core (smaller $Q_{C}$). The one-dimensional free energy profile also demonstrates switching of the relative stabilities of the two conformational states with folding/unfolding of Cter (Fig. 2c). These results demonstrate that Cter acts as a latch regulating the conformational states of the CD44-core, which is in line with previous experimental observation that appropriate conformational plasticity of CD44 is essential for stable cell rolling under fluid shear stress \cite{suzuki2015mechanical}. Because the T and R conformations have different HA binding-strengths, conformational switching modifies the CD44-HA interactions. As shown in Fig. 2d, when the CD44-core switches from the T conformation to the R conformation, the distribution of the ligand binding energy $E_{bind}$ is shifted, with the mean value increasing by $\sim$2.0 kcal/mol.

We further calculated the free energy landscape along the reaction coordinates $\chi$ and binding energy $E_{bind}$ (which characterizes the ligand dissociation). The conformations with $E_{bind} \geq $-0.0 kcal/mol is considered as dissociated state. One can see that after the CD44-core switches from T to R state, which is coupled to the unfolding of the Cter, the free energy barrier of ligand dissociation increases from $\sim$6.0 kcal/mol (megenta arrow) to $\sim$9.0 kcal/mol (blue arrow) (Fig. 2e), demonstrating the tight interplay between the folding state of the force-sensing Cter and the binding strength of ligand HA.

The above results suggest that unlocking of the force-sensing C-terminal latch under tensile forces tends to trigger the conformational switching of the CD44-core to the high binding-strength state via allosteric coupling, which can then contribute to the observed catch-bond effect. These findings directly demonstrate the crucial role of allosteric coupling, which is in line with the allosteric mechanism implicated in a number of previous catch-slip models \cite{thomas2006catch,thomas2008biophysics,thomas2008catch,sokurenko2008catch,sauer2016catch,evans2004mechanical,barsegov2005dynamics, phan2006remodeling,lou2007structure}.

\subsection{Tensile force induced repartitioning of parallel dissociation pathways}
To further understand the unusual triphasic force-response profile, we constructed the pathways of CD44-HA dissociation using MD trajectories under different pulling forces. The pathways were assigned based on the conformational state of the CD44-core (T or R), folding state of Cter (folded (F), detached (D), or partially folded intermediate (I)), and binding state of HA (bound (B) or unbound (U)) (Supplementary Text, Figs. S3 and S4). For example, ``TFB'' indicates that the CD44-core adopts the low-affinity T conformation, Cter is fully folded, and HA is bound. Strikingly, depending on the applied tensile force, we can observe three major dissociation pathways: 1) partial unfolding of Cter and T to R switching of the CD44-core first occur simultaneously, followed by full unfolding of Cter and then rupture of the CD44-HA bond from the high binding-strength R conformation (Fig. 3a-d); 2) conformational switching from the T to R state occurs simultaneously with full unfolding of Cter, and the CD44-HA bond ruptures from the high binding-strength R state (Fig. 3c, Fig. S4b); and 3) CD44-HA rupture occurs directly from the low binding-strength T state with Cter well folded (Fig. 3c, Fig. S4c). Both pathways 1 and 2 contribute to long-lifetime events.

Because the structure of the Cter latch, which is the key force-sensing element, can modulate the relative stabilities of the T and R conformations (Fig. 2b,c), the time scale of the conformational switching and the CD44-HA dissociation pathways are expected to be regulated by the strength of the applied tensile forces. In the small force range ($<5.2$ pN), the time scale of conformational switching is insensitive to the force strength (Fig. 3e), and HA tends to dissociate from the low binding-strength T conformation. Consequently, CD44-HA dissociation dominantly occurs following the pathway 3 (Fig. 3g). In this case, the major effect of the applied tensile forces is to reduce the barrier height of the CD44-HA rupture step, as described by the classic Bell law, demonstrating slip-bond behavior \cite{bell1978models}. With a further increase in the tensile force (5.2-12.1 pN), the T to R conformational switching becomes increasingly accelerated due to the force-induced unfolding of the force-sensing Cter latch (Fig. 3e) such that more CD44-HA rupture events occur in the high binding-strength R conformation following the pathways 1 and 2 (Fig. 3g). The increased populations of the high binding-strength pathways lead to elongation of the average lifetime of the CD44-HA bond, exhibiting the catch-bond effect. It is worth noting that the CD44-core may switch back to the low binding-strenth T state after the T to R conformational transition before ligand dissociation (Fig.
S5). Such T$\rightarrow$R$\rightarrow$T transition loop does not affect the relative populations of the slow and fast pathways, and therefore was omitted from the trajectory in constructing the pathways shown in Fig. 3c. With the increasing of tensile force, the reversal R to T transition becomes slower (Fig. 3e), which also tends to increase the probabilities of the high binding-strength pathways. At forces larger than a critical value ($\sim$12.1 pN), the relative population of the pathways 1 and 2 reached its maximum, and becomes less sensitive to the increasing of the tensile force. Consequently, the kinetics of the CD44-HA rupture recover to the slip-bond behavior. These results clearly demonstrate a force-induced multi-pathway repartitioning mechanism of the above observed slip-catch-slip triphasic pattern in the force-response behavior of the CD44-mediated adhesion bond. 

Due to the coexistence of multiple dissociation pathways with different rates at relatively large tensile forces, the observed survival probability deviates from single-exponent trend and show piecewise linear decays with distinct slopes in the semi-log plots (Fig. 3f, Fig. S6). The fast decay phase and slow decay phase are dominantly contributed by the dissociation events following the fast pathway 3 (Fig. 3f, red dots) and the slow pathways 1 and 2 (Fig. 3f, blue dots), respectively. The lifetime distributions for the dissociation events following the above two sets of pathways can be separately fitted by single-exponential functions with different time constants (Fig. S6). Such results are consistent with the force-induced multi-pathway repartitioning mechanism discussed above.

\begin{figure*}[hbt!]
\centering
\includegraphics[width=0.9\linewidth]{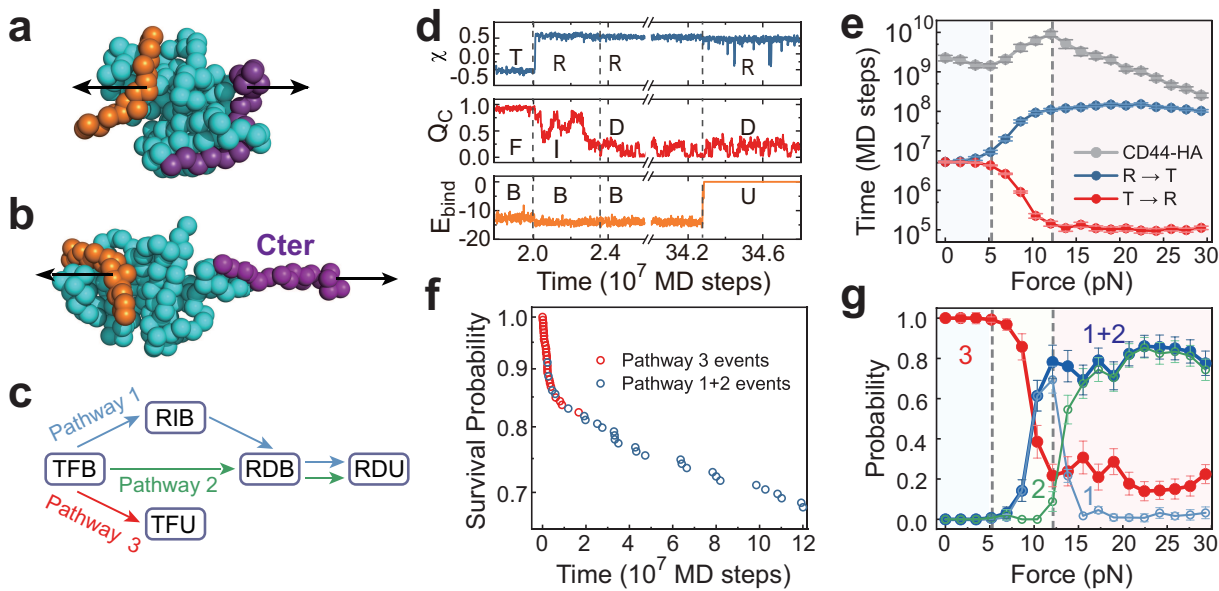}
\caption{Tensile force-induced kinetic repartitioning of the CD44-HA rupture pathways. (a,b) Three-dimensional structure showing representative snapshots of the pulling simulations in the initial TFB state (a) and the later RDB state (b). (c) Schematic showing the three major pathways of CD44-HA rupture. (d) Representative trajectory of the pathway 1 showing the $\chi$ value (upper), $Q_C$ score (middle), and HA binding energy (bottom, the unit is kcal/mol) as a function of time. (e) Average time for the transition from the T conformation to the R conformation and its reversal process as a function of pulling force. For comparison, the average lifetime of the CD44-HA bond as a function of the applied tensile force is also shown (the same as that shown in Fig. 1f). (f) Survival probability of the CD44-HA bond as a function of simulation time under the tensile force of 24.2 pN plotted by an event-by-event manner. The dissociation events following the fast and slow pathways were colored red and blue, respectively. (g) Probabilities of the three major pathways as a function of the pulling force. The results shown in (e-g) were calculated based on at least 192 independent simulations for each tensile force. The error bars were estimated based on bootstrap method.}
\label{fig3}
\end{figure*}

\subsection{Bidirectional allostery mechanism of catch-bond effect}
The above results show that unlocking of the C-terminal latch tends to trigger the conformational switch of the CD44-core to the high binding-strength R state, which therefore contributes to enhanced CD44-HA binding. However, before Cter unfolding, the CD44-core mostly stays in the low binding-strength T conformation as discussed above, and applying tensile force also tends to promote rupture of such low strength CD44-HA bond. Apparently, developing the catch-bond effect requires that there are a sufficient number of events in which force-induced unfolding of the Cter occurs prior to CD44-HA rupture. Consequently, a natural question that arises is what strategy the CD44 uses to meet the above required specific kinetic control over a wide range of tensile forces, which is the key aspect for understanding the molecular mechanism of the catch-bond effect but still remains unresolved. 

To answer the above question, we characterized the force response behavior of the C-terminal latch, which is the force-sensing element of the CD44. Firstly, we performed constant-velocity pulling simulations with the pulling force applied between the CD44-core and the C-terminal residue and extracted the unfolding forces of the Cter latch (Fig. 4a). By repeating multiple simulations with and without HA binding, we calculated the distributions of the unfolding forces. One can see that HA binding dramatically modifies the force response behavior of the C-terminal latch (Fig. 4b). Without HA binding, the C-terminal latch shows high mechanical stability with the mean value of the unfolding force of $\sim$17.0 pN, which is much higher than the typical force range showing catch-bond effect (5.2pN-12.1pN, Fig. 1f). In contrast, after HA binding, the mean value of the unfolding force becomes $\sim$9.0 pN, which is within the force range of the catch-bond phase. Similar results can be observed when different force acting points were used in the pulling simulations (Fig. S7).

In the above constant-velocity pulling simulations, the used pulling velocity needs to be very high in order to observe a successful unfolding event within reasonable simulation time, which tends to overestimate the unfolding force. Therefore, we also performed constant-force pulling simulations at different tensile forces and calculated the free energy profiles. The pulling force at which the folded state and unfolded state have equal population is defined as unfolding force. The results show again that HA binding significantly reduces the unfolding force (Fig. 4c,d and Fig. S8). In the absence of HA, the unfolding force is between 10.4 pN and 12.1 pN. In the presence of HA, the unfolding force becomes $\sim$ 6.9 pN. The free energy profile along the reaction coordinate $Q_{C}$ at zero force load also showed that the free energy barrier height for Cter unfolding was significantly lowered upon HA binding (Fig. S3). All these results suggest that the binding of the ligand HA significantly decreases the stability and unfolding force of the C-terminal latch, although the HA binding site is spatially distant from the Cter, demonstrating typical allosteric coupling behavior. Therefore, the above results indicate that CD44 is able to utilize the allosteric coupling from the HA binding site to the force-sensing element Cter to promote the detachment of the otherwise stably folded C-terminal latch with a much lowered threshold force. Consequently, it may enhance  the  force  sensitivity of the C-terminal latch and enable the force-induced conformational switching to occur before the bond rupture from the low binding-strength state which is a basic prerequisite for catch-bond formation.  

The above results suggest a bidirectional allosteric coupling mechanism of force-enhanced cell adhesion. In this mechanism, HA binding allosterically promotes the mechanical unfolding of the distant force-sensing Cter before bond rupture (i.e., forward allostery), which in turn activates the CD44-core and leads to force-enhanced CD44-HA binding via backward allostery (Fig. 4e). Such bidirectional allosteric coupling enables the catch-bond effect to occur under a wide range of tensile forces and therefore may favor efficient rolling adhesion of cells.

To further demonstrate the role of bidirectional allostery for catch-bond formation, we constructed a minimalist kinetic model (Fig. 4f). The unfolding rate of Cter under tensile force F can be estimated as 
\begin{equation}
k_u(F)=k_1^0 e^{\beta F\Delta x_1}. 
\end{equation}
Here, $k_1^0$ represents the unfolding rate of Cter at zero-force without considering the forward allostery, $\Delta x_1$ represents the transition distance for unfolding, and $\beta = 1/k_BT$. When the effect of forward allostery is considered, the barrier height for the Cter unfolding is decreased by $\Delta \Delta G$. According to the simulation result (Fig. S3), $\Delta \Delta G \sim$ 2kcal/mol. The unfolding rate is then written as following
\begin{equation}
k_u(F)=k_1^0 exp(\beta\Delta \Delta G+\beta F\Delta x_1). 
\end{equation}
The Cter unfolding is reversible with the folding rate given by $k_2^0 e^{-\beta F\Delta x_2}$. HA may dissociate either from the Cter-folded state or Cter-unfolded state, with the rates being roughly estimated by $k_3^0 e^{-\beta F\Delta x_3}$ and $k_4^0 e^{-\beta F\Delta x_4}$, respectively. Here $k_2^0$, $k_3^0$, $k_4^0$, $\Delta x_2$, $\Delta x_3$, and $\Delta x_4$ represent the corresponding rates and transition distance at zero-force. The backward allostery, as implicated in most catch-bond models, requires $k_3^0 > k_4^0$, such that unfolding of the force sensor Cter elongates the CD44-HA lifetime. By appropriately optimizing the parameters (Tab. S1), the above bidirectional model can reasonably describe the slip-catch-slip triphasic behavior observed in the simulations (Fig. 4g, orange). In comparison, when the forward allostery is not considered ($\Delta\Delta G=0$), the catch-bond effect becomes much weaker (Fig. 4g, blue). Such results clearly demonstrates that the forward allostery plays crucial role for the catch-bond formation, supporting the bidirectional allostery mechanism.

\begin{figure*}[hbt!]
\centering
\includegraphics[width=0.9\linewidth]{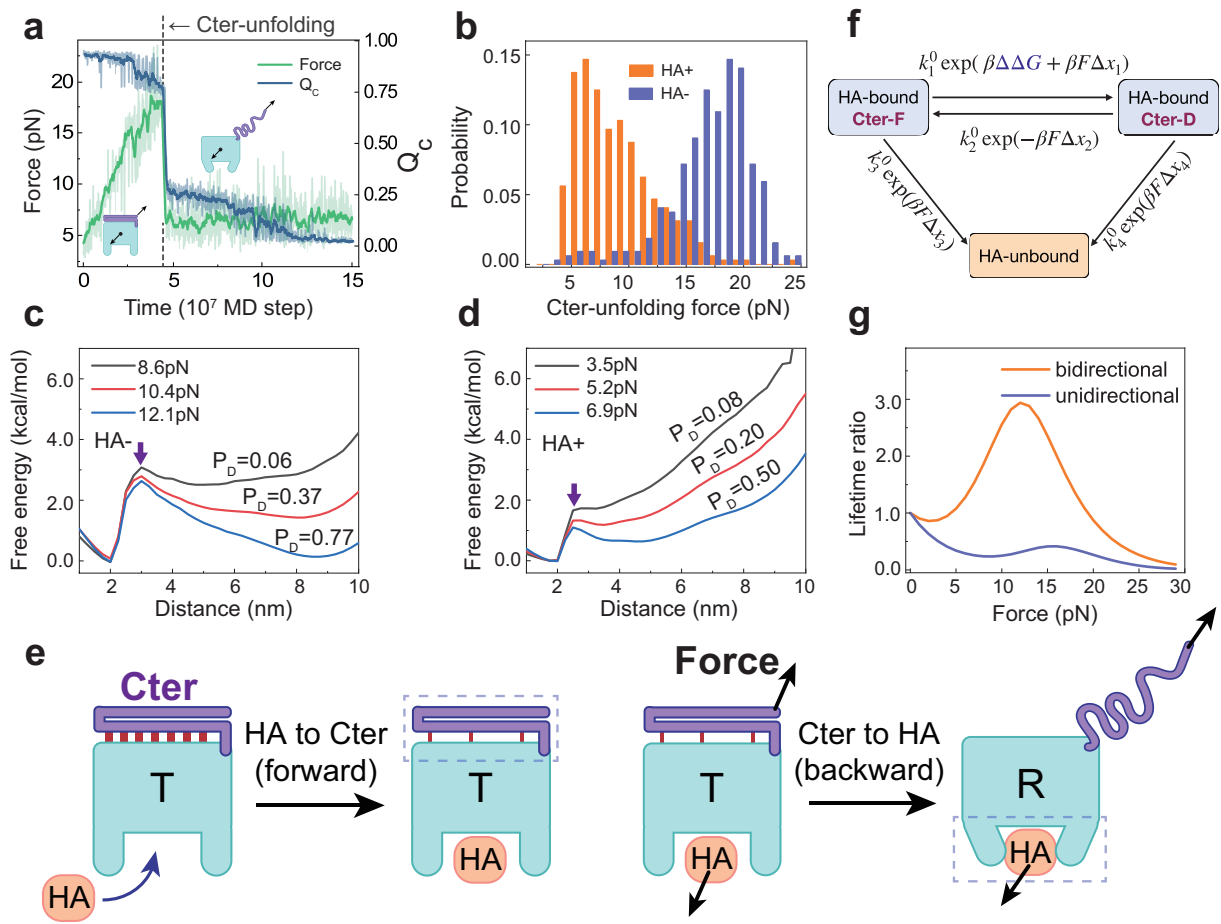}
\caption{HA binding-induced allosteric coupling promotes unlocking of the force-sensing Cter latch. 
 (a) Force (green) and $Q_C$ (blue) as a function of time for a representative trajectory of constant velocity pulling simulations. The pulling force was applied between the center of mass of the CD44-core and the C-terminal residue. The unfolding event was indicated by the dashed line, from which one can extract the unfolding force. (b) Distributions of unfolding force of Cter with (orange) and without (blue) binding of HA. (c,d)One-dimensional free energy profiles along the reaction coordinate defined by the distance between the start and the end of the Cter at different pulling forces without (c) and with (d) the presence of HA. The corresponding populations of the Cter-unfolded conformations ($P_D$) were also shown. (e) Cartoon diagram showing the bidirectional allosteric coupling mechanism of the catch-bond effect. (f) A kinetic model of catch-bond considering the bidirectional allostery effect. Cter-F and Cter-D represent the conformational states with Cter folded and unfolded, respectively. The parameters in the kinetic model were optimized by fitting the simulated lifetime at different tensile forces (Fig. 1f). The value of $\Delta\Delta G$, which describes the effect of forward allostery, was set to 2.0 kcal/mol. (g) Average lifetime of the CD44-HA bond as a function of the applied tensile force calculated by the kinetic model with (orange) and without (blue, $\Delta\Delta G=0$) considering the forward allostery effect.}
\label{fig4}
\end{figure*}

As a complementary characterization of the bidirectional allostery mechanism, we further conducted two sets of all-atom MD simulations, i.e., ``ligand-bound'' simulations and ``ligand-free'' simulations. In order to avoid introducing new force field parameters, the ligand HA was considered implicitly in the ``ligand-bound'' simulations (Appendix D, Supplementary Text). Initiating from the ordered structure with the Cter being well folded, we conducted five independent simulations for each case under force-free condition. We then performed energy decomposition analysis to investigate the allosteric coupling\cite{gmx_mmpbsa}. The calculated contact energies between the residues of the Cter and the CD44-core showed that binding of HA significantly weakens the key interactions stabilizing the Cter (Fig. 5b). Particularly, the interface interactions involving the Tyr161 of the Cter, which has the most prominent contribution to the interface interactions between the Cter and the CD44-core, were much weakened upon ligand binding (Fig. 5c-f). For the ligand-free case, the Tyr161 forms stable contacts with the CD44-core with the average contacting energy of $\sim$ -13.5 kcal/mol. For the ligand-bound case, the Tyr161 hops between a strongly contacting state ($\sim$ -13.5 kcal/mol) and a weakly contacting state ($\sim$ -7.9 kcal/mol). Apparently, such ligand binding induced destabilization of the interface interactions between the CD44-core and Cter would reduce the threshold force for the Cter unfolding. All these all-atom MD simulation results and the energetic decomposition analysis suggest that the allosteric coupling from the HA binding site to the Cter is significant, further supporting the bidirectional allostery mechanism. We also conducted  all-atom  MD simulations and energy decomposition analysis for the PSGL-1 receptor P-selectin, which is another key receptor mediating rolling adhesion of cells along vessel walls during the inflammatory response
\cite{marshall2003direct,phan2006remodeling,evans2004mechanical,evans2007forces}, and similar results were observed (Fig. S9).

\begin{figure*}[hbt!]
\centering
\includegraphics[width=0.9\linewidth]{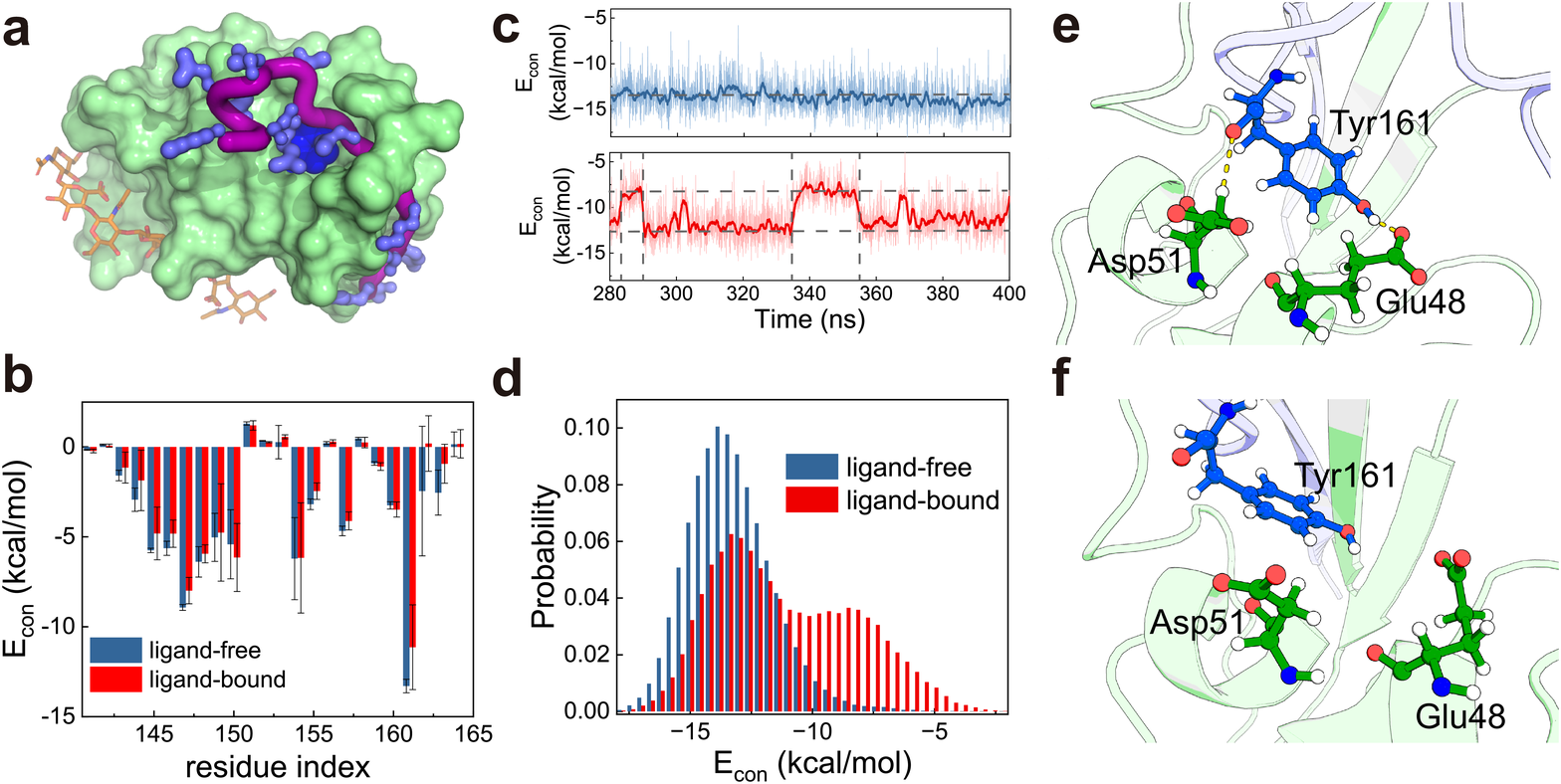}
\caption{Effect of HA binding on the stability of Cter based on all-atom MD simulations and energy decomposition analysis. (a) Three-dimensional structure of CD44 showing the interface between the CD44-core (light green) and the Cter (purple tube). The sidechains of the Cter were shown explicitly. (b) Contact energies between the CD44-core and the Cter decomposed onto the residues of the Cter for the ligand-free and ligand-bound simulations. (c) Representative trajectories showing the contact energy involving the Tyr161, which is shown by dark blue spheres in (a), for the ligand-free (blue) and ligand-bound (red) simulations. (d) Probability distributions of the contact energy between the Tyr161 and the CD44-core for the two sets of simulations. (e,f) Representative structures showing different contacting modes of the Tyr161 from the ligand-free (e) and ligand-bound (f) simulations.}
\label{fig5}
\end{figure*}

\subsection{Role of receptor conformational equilibrium in shaping the catch-bond}\label{subsec2.5}

In the above discussions, the protein model was calibrated to satisfy the experimental restraints on the conformational equilibrium, therefore corresponding to wild-type (WT) CD44 (Supplementary Text). The conformational equilibrium is related to the energy landscape dictating the conformational switching of CD44-core between the T and R states, and is controlled by the global energy gap parameter $\Delta V$ of the double-basin energy function given by Eq. 2 (Fig. 6a, Appendix A). To demonstrate the crucial role of appropriate conformational equilibrium in the observed catch-bond effect, we also conducted simulations by designing a number of CD44-mutant models, in which the protein conformational equilibrium was altered by tuning the parameter $\Delta V$. For the CD44-mutant models with an extremely T- or R-biased conformational equilibrium, the lifetime of the CD44-HA bond monotonically decreases with increasing tensile force (Fig. 6b), lacking the catch-bond effect. Nevertheless, one can observe obvious catch-bond effect for the parameter range around the WT model (Fig. 6b, c). These CD44 models exhibiting catch-bond effect have two key features, which may represent the essential thermodynamic and kinetic requirements for the receptor protein to shape a catch-bond. First, both the T and R conformations are accessible with significant probabilities in the zero force case, and applying a tensile force can modulate the conformational distribution through force-induced unfolding of the Cter(Fig. 6d). Therefore, the relative stability between the two conformational states should not be too large, such that the conformational dynamics of the CD44-core can be sensitively modulated by the Cter unfolding. Second, the time for the reverse conformational switching from the R to T state can be significantly prolonged by applying a tensile force due to force-induced Cter unfolding, such that the receptor has higher probability to be locked in the high binding-strength conformation before bond rupture (Fig. 6e), demonstrating the crucial role of kinetic effect of tensile forces for catch-bond formation.

The above results demonstrate that the CD44 models with appropriate conformational equilibrium tend to exhibit more prominent catch-bond effect, possibly because these protein models can better satisfy the above thermodynamic and kinetic requirements for bidirectional allosteric coupling.
These results may suggest that WT CD44 has well evolved to optimize the conformational equilibrium and thereby develop the catch-bond effect, which is essential for its biological function to mediate rolling adhesion of cells. Such a tight interplay between the appropriate conformational equilibrium and force-response behavior of adhesion bonds has significant implications for drug design targeting cell adhesion receptors. 

At physiologically relevant range of shear stresses, the circulating tumor cells or leukocytes may behavior diversely\cite{richter2012interaction,gal2003role}. Part of the cells undergo rolling adhesion, and other cells may undergo firm adhesion or directly flow away with the fluids. The relative importance of these different outcomes varies with the cell lines and the shear stress. In a previous experimental work by Suzuki and coworkers \cite{suzuki2015mechanical}, two CD44 mutants respectively over-stabilizing the high binding-strength conformation (R) and the low binding-strength conformation (T) were introduced. Their results showed that for the mutant biasing to high binding-strength conformation (Y161A), the rolling of the cells is much less efficient. Whereas for the mutant biasing to the low binding-strength conformation (by introducing a disulfide bond between T47 and N164), the adhesion becomes much weaker and rolling adhesion is rare. These mutagenesis studies suggested that mutations altering the conformational equilibrium of the CD44 may affect the biological function, which supports the key role of conformational equilibrium of receptor protein for the catch-bond formation revealed above. 

In addition to the conformational equilibrium of receptor protein, the force response behavior of the CD44-HA bond could also be modulated by the variations of the ligand HA. In this work, the key feature of HA as a native ligand of CD44 is considered by appropriately assigning the interaction strength with CD44. The coefficient $\lambda\epsilon_{LP}$ in Eq. A2 controls the interaction strength between HA and CD44. In order to investigate the effect of the variations in HA on the observed force-response behavior of the CD44-HA bond, we performed additional simulations with varied interaction strengths $\lambda\epsilon_{LP}$. The value of $\lambda$ less (larger) than 1.0 represents that the varied HA has weaker (higher) binding strength with the CD44, by which we can effectively model the effect of variations in HA on the force-response behavior of CD44-HA bond. The results showed that for wide range of interaction strengths, we can observe the catch-bond effect. However, when CD44-HA interactions become extremely weak or strong, applying tensile force does not enhance the lifetime of the CD44-HA bond (Fig. S10), and therefore
the catch-bond effect becomes weak.

\begin{figure*}[hbt!]
\centering
\includegraphics[width=1.0\linewidth]{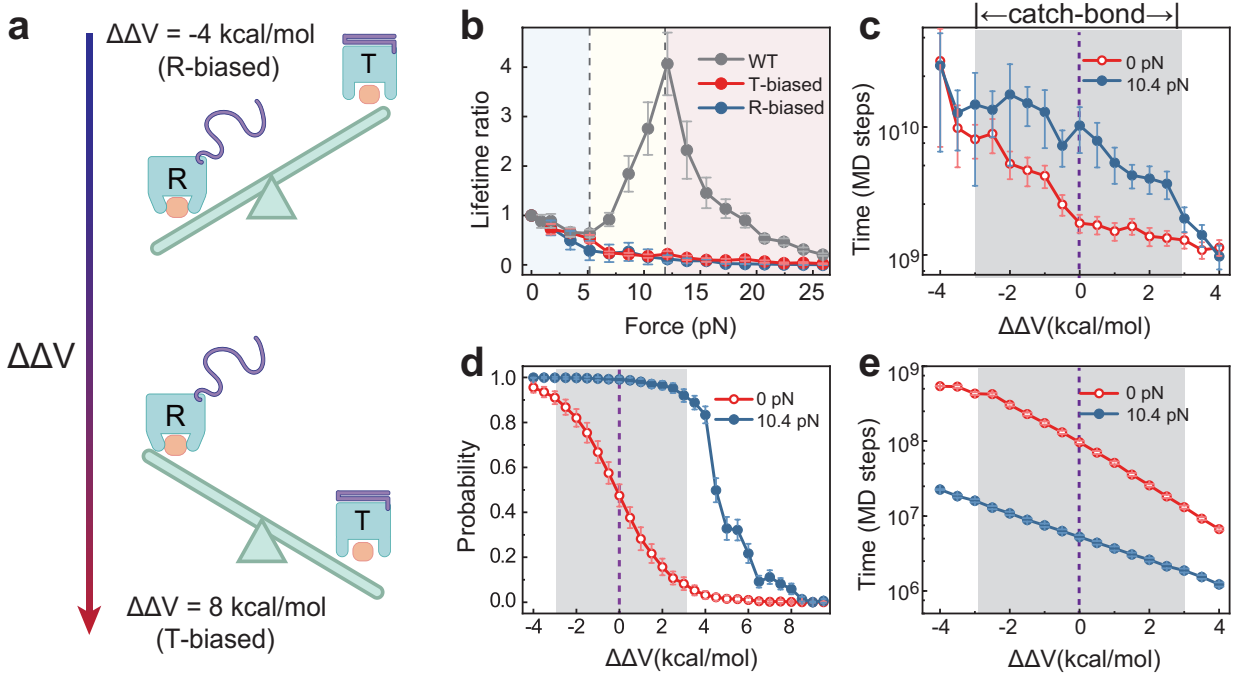}
\caption{Interplay between conformational equilibrium of receptor protein and catch-bond formation. (a) Cartoon illustration for the construction of the CD44-mutant models with different conformational equilibrium by modifying the energy gap parameter ($\Delta\Delta V$ represents the change of $\Delta V$) in the dynamic energy landscape model. (b) Average lifetime of CD44-HA bond as a function of tensile force for two mutant models of CD44 with the conformational equilibrium over-biased to T state (red) and R state (blue), respectively. For comparison, the results for the WT CD44 were also shown (gray, the same as that shown in Fig. 1f). (c) Average lifetime of CD44-HA bond as a function of energy gap change $\Delta\Delta V$ with zero force load (red) and with a tensile force of 10.4 pN (blue). Positive and negative $\Delta\Delta V$ values correspond to the T-biased and R-biased CD44 mutant models, respectively, relative to the WT model ($\Delta\Delta V=0$, purple dash line). (d) Probability for the CD44-core to stay at R state as a function of $\Delta\Delta V$ with zero force load (red) and with a tensile force of 10.4 pN (blue). (e) The average time of the conformational switching from the high binding-strength R conformation to the low binding-strength T conformation as a function of $\Delta\Delta V$ with zero force load (red) and with a tensile force of 10.4 pN (blue). The error bars were estimated based on at least 192 independent
simulations using bootstrap method.}
\label{fig6}
\end{figure*}

\section{Discussions and conclusion}\label{discussion}
The catch-bond effect is widely used by cells to achieve efficient adhesion under mechanical stress and plays critically important roles in many biological and pathological processes. By constructing a residue-resolved dynamic energy landscape model supported by available experimental restraints, we could simulate the full molecular event of receptor-ligand dissociation under tensile forces and study the underlying molecular mechanism of the catch-bond effect. We carried out MD simulations for the CD44-HA adhesion bond with tensile forces, which is involved in rolling adhesion of many circulating tumor cells, leukocytes, and other cells under fluid shear stress. The molecular simulation results revealed a bidirectional allosteric coupling mechanism for the catch-bond effect, whereby HA binding allosterically facilitates detachment of the otherwise stably folded Cter latch with a much reduced threshold force. The detachment of Cter in turn stabilizes the high binding-strength conformation of the CD44-core through backward allosteric coupling, which therefore causes a prolonged lifetime of the ligand-bound state, leading to the catch-bond effect. Such bidirectional allostery mechanism provides a solution for receptor protein to satisfy the basic prerequisite for catch-bond formation, namely, force-induced remodeling of ligand binding interface occurs prior to bond rupture. It is worth emphasizing that the above bidirectional allosteric coupling mechanism does not rely on rapid equilibrium between the high binding-strength and low binding-strength conformations. Instead, the relative stability and the timescales for the transitions between the two conformations are more important factors contributing to the catch-bond effect. In addition, the results of this work revealed a force-induced kinetic repartitioning and its key role in the unusual force-response behavior of CD44-HA bond. The rupture of CD44-HA bond can follow multiple parallel pathways. An increase in the tensile force promotes Cter unfolding and the population of the high binding-strength conformation through the above bidirectional allosteric coupling mechanism, which therefore favors the long-lifetime CD44-HA dissociation pathways, leading to reinforcement of the CD44-HA bond with increasing tensile force. 

These findings provide mechanistic insights into the molecular mechanism of immunity, infection, tumor metastasis, and other biological processes. The mean shear stress in blood vessel or lymphatics ranges from less than 0.1 dyne/cm$^2$ to tens of dyne/cm$^2$, depending on the locations \cite{oyre1997vivo,dixon2006lymph}. \emph{In vitro} experimental measurements showed that the shear force tends to reinforce the interactions between HA and leukemic cell line expressing CD44 for the shear stresses at or higher than 0.2 dyne/cm$^2$, with the maximum adhesion strength being observed at $\sim$1.0 dyne/cm$^2$\cite{christophis2011shear}. Therefore, the shear stress range showing enhanced rolling adhesion is relevant to physiological conditions. Mapping to the range of tensile force for a single receptor-ligand bond, the CD44-HA bond shows typical catch-bond effect at the tensile force from several piconewton to tens of piconewton \cite{bano2016single}, which covers the force range investigated in this work.

This study suggests that the catch-bond behavior in the force responses of cell adhesion involves both the allosteric feature of the receptor and force-induced kinetic repartitioning between parallel dissociation pathways. Interestingly, both the two-pathway mechanism and the allostery mechanism have been proposed to explain the observed catch-bond effect for a number of receptor proteins \cite{thomas2006catch,sauer2016catch,evans2004mechanical,barsegov2005dynamics,pereverzev2005two,phan2006remodeling,waldron2009transmission}. Here, we showed that these two different mechanisms are not mutually exclusive. Instead, they emphasize different aspects of the dynamic processes of receptor-ligand bond rupture, and both are crucial for shaping the observed unusual force responses in cell rolling adhesion. Notably, although allosteric coupling is intrinsically bidirectional, the classic allosteric mechanism of the catch-bond effect often emphasizes the coupling from the force-sensing element to the ligand binding site. The bidirectional allosteric coupling mechanism revealed in this work demonstrates the crucial role of allosteric coupling from the ligand binding site to the force-sensing Cter latch by promoting Cter unfolding, which is essential for meeting the specific kinetic control required for the catch bond. We propose that such bidirectional allosteric coupling may be general for other receptor proteins mediating cell adhesion. In addition, the dynamic energy landscape model constructed in this work provides a useful framework for describing the molecular events involved in cell adhesion under tensile forces.

The bidirectional allostery mechanism and force-induced kinetic repartitioning revealed in this work provide the global thermodynamic and kinetic principle shaping the catch-bond effect. However, the local atomic-level molecular event occurred in the ligand binding pocket may also contribute to the observed force-response behavior of receptor-ligand bond, as suggested by the deformation model and sliding-rebinding model proposed in early studies\cite{pereverzev2006force,lou2007structure,prezhdo2009theoretical}. It is worth noting that the atomistic details of the binding site during the ligand dissociation cannot be explicitly captured in the current molecular simulations with the dynamic energy landscape model, because of the coarse-graining for both the ligand and receptor. Therefore, the bidirectional allostery mechanism does not exclude possible contributions from other pathways arising from atomistic-level arrangement of binding site. It is interesting in future studies to further investigate the interplay between such atomic-level local molecular event and the above global thermodynamic and kinetic principle in the catch-bond formation by developing atom-resolved computational model.

In this work, in addition to reproducing the catch-slip biphasic transitions typically observed in previous experimental works, the MD simulations revealed a triphasic slip-catch-slip pattern of the force responses which is featured by an additional slip-bond phase at the low force end (0-5.2pN). As previous single-molecule experimental works were often conducted at tensile forces larger than 5.0 pN, this initial slip-bond phase has not been reported for the CD44-mediated adhesion bond. Interestingly, previous experimental works using a flow chamber and a biomembrane force probe observed a triphasic force dependence for E-selectin mediated cell adhesion \cite{wayman2010triphasic}, which is beyond the physical models developed for biphasic catch-slip bonds but can be well captured by the above dynamic energy landscape model. Performing single-molecule measurements in the small force region to test the prediction of the slip-catch-slip triphasic pattern of the force dependence for the CD44-mediated adhesion bond would be interesting in future studies.

\begin{acknowledgments}
This work was supported by the National Natural Science Foundation of China (Nos. 11974173 and 11934008) and the HPC Center of Nanjing University.
\end{acknowledgments}

\appendix
\section{Dynamic energy landscape model of cell adhesion}
In this work, we used the atomic interaction-based coarse-grained model with flexible local potential (AICG2+) \cite{li2014energy}, which has a structure-based energy function and was 
constructed based on the energy landscape theory of protein folding assuming a funnel-like energy landscapes with minimal energetic frustration \cite{clementi2000topological,onuchic1997theory}, to describe the conformational motions of the CD44-core around native structures and Cter folding. The parameters were determined by using a multiscale strategy\cite{li2011frustration,li2014energy}. In this model, each residue was represented by a spherical bead located at the $C_\alpha$ position (Fig. S1), and its energy function is given by
\begin{equation}
V(\overrightarrow{\mathbfit{x}}\mid\overrightarrow{\mathbfit{x}}_0)=V_{bond}+V_{loc}^{FLP}+V^{SB}+V_{exv}+V_{ele}
\end{equation}
In the above formula, $\overrightarrow{\mathbfit{x}}$ and $\overrightarrow{\mathbfit{x}}_0$ represent the coordinates of the coarse-grained beads in a given protein structure and the native structure, respectively. $V_{bond}$ describes the covalent connectivity between the consecutive beads. $V_{loc}^{FLP}$ is the local potential describing the chain flexibility and secondary structure propensity, which was extracted by statistical survey of the coiled library \cite{terakawa2011multiscale,li2012energy}.  $V^{SB}$ represents the structured-based potential\cite{clementi2000topological}, which shapes up a funneled energy landscape driving the folding of the proteins to their native structures\cite{onuchic1997theory}. The $V_{exv}$ represents the excluded volume term. For the charged residue pairs, the electrostatic interaction term  $V_{ele}$ was applied, which is given by the Debye-H$\ddot{u}$ckel formula. More details of the AICG2+ energy function can be found in Ref. \cite{li2014energy}. Based on the above AICG2+ model, we further constructed the double-basin energy function given in Eq. 2 to describe the conformational switching between the T and R states \cite{okazaki2006multiple}, which is essential for efficient cell rolling adhesion. The terms $V(\overrightarrow{\mathbfit{x}}\mid\overrightarrow{\mathbfit{x}}^{T}_0)$ and $V(\overrightarrow{\mathbfit{x}}\mid\overrightarrow{\mathbfit{x}}^{R}_0)$ in Eq. 2 are given by the AICG2+ potentials centric to the T and R structures, respectively. The HA bound partially disordered structure (PDB code:2I83)\cite{takeda2006ligand} and the HA unbound ordered structure (PDB code:1UUH)\cite{teriete2004structure} determined in previous experimental works were used as the reference structures of the structure-based energy functions for the R and T conformations of the CD44-core, respectively. These reference native structures defined the minima of the two energy basins of the above double-basin energy function. In the double-basin energy function, the key parameter $\Delta V$ controls the relative stabilities (free energy difference) of the T and R conformations. As shown in previous experimental data \cite{ogino2010two,suzuki2015mechanical}, the T and R states are the dominant states in the absence and presence of HA, respectively. In this work, the $\Delta V$ value was set to reproduce the above qualitative trend observed experimentally. In the simulations, the optimized value of the parameter $\Delta V$ was fixed and independent of the tensile force. As a test of parameter sensitivity, we also conducted the simulations with varied $\Delta V$ values(Fig. 6). Although the realistic energy landscape should be more rugged, the above energy function with two-basin topography represents a reasonable simplification of a more complicated energy landscape. Such double-basin energy landscape has been successfully used in describing the allosteric coupling involved in a large number of protein systems, such as molecular motors, enzymes, transporters, and signal proteins in previous works\cite{yao2010drug,maragakis2005large,li2019overcoming,whitford2007conformational}. Similarly, the energy function of the C-terminal segment of the CD44 was described by the AICG2+ potential with the ordered structure being used as the reference native structure. Only the HA binding domain of the CD44 was included in the simulations. 

Because the coordinates of the HA atoms are missing in the solution structure of the HA bound human CD44\cite{takeda2006ligand}, we reconstructed the coordinates of the HA atoms based on the crystal structure of the HA bound murine CD44 by superimposing the atoms of the ligand binding residues \cite{banerji2007structures}. The reconstructed structure of the HA binding site was then used to build the AICG2+ potential for the HA binding. Each disaccharide unit of HA was represented by five beads located at the O5, N2, O3, C2, and C5 positions (Fig. S1). The bond length, bond angle, and dihedral angle formed by these coarse-grained beads were restrained to their corresponding values in the PDB structure by harmonic potentials following previous work \cite{yao2010drug}. The interactions between HA and CD44 were described by
\begin{equation}
V_{LP}=\lambda\epsilon_{LP}\sum_{ij}\left[5\left(\frac{r_0^{ij}}{r^{ij}}\right)^{12}-6\left(\frac{r_0^{ij}}{r^{ij}}\right)^{10}\right]+V_{exv}
\end{equation}
where the sum index $i$ runs over all the beads of HA, and the sum index $j$ runs over the residues of the ligand binding site of CD44. $r_{ij}$ and $r_{ij}^0$ are the distances between the HA bead $i$ and the residue $j$ of binding site in a given snapshot of the molecular simulations and in the HA bound structure. The coefficents $\lambda\epsilon_{LP}$ controls the binding strength of HA to CD44 (Supplementary Text).

As HA binding reshapes the intrinsic energy landscape of CD44, the final energy function dictating the conformational switching between the T and R states depends on the HA binding state, which therefore corresponds to a dynamic energy landscape model. In addition, the folding state of the Cter can also modulate the relative stability of the two conformational states. Applying tensile force may induce the unfolding of the Cter, which then results in force dependence of the conformational dynamics of the CD44-core and the lifetime of the CD44-HA bond. More details of the computational model are given in Supplementary Text\cite{guan2023SI}.

\section{Coarse-grained molecular simulations}
Molecular simulations were conducted based on Langevin dynamics with friction coefficient $\gamma=0.25\tau^{-1}$ and temperature $T=310.2~K$ using CafeMol3.0 software \cite{kenzaki2011cafemol}. The time step was set as 0.1$\tau$, with $\tau$ being the reduced time unit in CafeMol. We conducted at least 192 independent simulations for each tensile force. The simulations under different tensile forces were initiated from the ordered structure with HA bound, at which the CD44-core adopts a low-affinity T conformation and the C-terminal segment (Cter) is well folded. In all the simulations, only the hyaluronan-binding domain of CD44 was included\cite{ogino2010two}.

In investigating the effect of HA binding on the unfolding force of the C-terminal segment, we also conducted constant velocity pulling simulations and constant force pulling simulations in the presence and absence of ligand HA (Supplementary Text). We conducted 160 independent MD simulations lasting for $2 \times10^8$ MD steps with the HA bound and unbound, respectively, by which we can extract the unfolding forces and the free energy profiles. 

\section{Reaction coordinates}
We introduced the reaction coordinates $\chi$ and $Q_{C}$ to characterize allosteric coupling. The reaction coordinate $Q_C$ describes the fraction of the formed native contacts in a given structure, which is defined by
\begin{equation}
Q_C=\frac{1}{N_{C}^{nat}}\sum_{kl}\frac{1}{1+\exp[-\beta(\lambda-r_{kl}/r_{kl}^0)]}
\end{equation}
where $N_{C}^{nat}$ is the number of native contacts between the CD44-core and Cter. The sum index runs over all the residue pairs forming the native contacts. $r_{kl}$ and $r_{kl}^0$ are the distances of the residue pair in a given snapshot and in the native structure, respectively. The $\beta$ and $\lambda$ are two parameters with the values of $5.0$ and $1.2$, respectively. We also introduced the binding energy $E_{bind}$ to describe the interaction strength between the CD44-core and HA and assign the binding states. 

The reaction coordinate $\chi$ describes the conformational change of the CD44-core between the T and R conformations, and is given by \cite{okazaki2006multiple}
\begin{equation}
\exp\left(2\chi\right)=\frac{V\left(\overrightarrow{\mathbfit{x}}\mid\overrightarrow{\mathbfit{x}}^{R}_0\right)-V_{DB}\left(\overrightarrow{\mathbfit{x}}\right)}{V\left(\overrightarrow{\mathbfit{x}}\mid\overrightarrow{\mathbfit{x}}^{T}_0\right)-V_{DB}\left(\overrightarrow{\mathbfit{x}}\right)}
\end{equation}
Negative and positive values of the $\chi$ represent the T and R conformations, respectively. 

\section{All-atom molecular dynamics simulations}
All-atom MD simulations for CD44 and P-selectin were conducted by using Gromacs 2019.6 combined with PLUMED\cite{tribello_plumed_2014}. We conducted two sets of all-atom MD simulations, i.e., ``ligand-bound'' simulations and ``ligand-free'' simulations. In the ``ligand-bound'' simulations, in order to avoid introducing new force field parameters for the ligand HA, we performed the simulations with implicit representation of ligand HA. The effect of HA binding was modelled by restraining the residues at the HA binding site to the corresponding conformation at the HA-bound structure (pdb entry: 2I83) using a harmonic biasing potential. By this way, we can effectively simulate the effect of HA binding without explicitly including HA in the simulations. As a control, we also performed ``ligand-free'' simulations, in which the residues at the HA binding site being restrained to the corresponding conformation at the ligand-free apo structure (pdb entry: 1UUH). All the simulations were initiated from the ordered structure with the Cter being well folded. Five independent simulations with the length of 400 ns were conducted for each case at 300 K and 1 atm under force-free condition. Then, we calculated the contact energies between the residues of the Cter and the CD44-core by performing energy decomposition analysis with the gmx$\_$MMPBSA tool\cite{gmx_mmpbsa}. We also performed the similar all-atom MD simulations and energy decomposition analysis for another receptor protein P-selectin mediating cell adhesion. Details of all-atom MD simulations were given in Supplementary Text.

\bibliography{main}

\end{document}